\documentclass[12pt,preprint2]{aastex}

\newcommand{\ch}{{\it Chandra }}
\newcommand{\xmm}{{\it XMM-Newton}}
\newcommand{\fu}{{\it FUSE }}
\newcommand{\ulg}{ULIRGs}

\newcommand{\ergs} {erg s$^{-1}$}

\newcommand{\afe}{$\alpha/\rm{Fe}$}

\newcommand{\ang}{\rm{\AA} }
\newcommand{\vv}{VV 114}
\newcommand{\ks}{\rm{km/s}}
\newcommand{\lfir}{$\rm{L_{FIR}}$}
\newcommand{\ovi}{\ion{O}{6}}

\shorttitle{FUSE \& X-ray~Observations of VV 114}
\shortauthors{J Grimes et al.}

\begin{document}

\title{Far-Ultraviolet \& X-ray~Observations of \vv: Feedback in a Local Analog to 
Lyman Break Galaxies\\
{\it Accepted for Publication in ApJ}}

\author{J. P. Grimes\altaffilmark{1}, T. Heckman\altaffilmark{1}, C. Hoopes\altaffilmark{1}, 
D. Strickland\altaffilmark{1}, A. Aloisi\altaffilmark{2}, G. Meurer\altaffilmark{1}, and A. Ptak\altaffilmark{1}}
\altaffiltext{1}{Center for Astrophysical Sciences, Johns Hopkins University,
    3400 N. Charles St, Baltimore, MD 21218; jgrimes@pha.jhu.edu, heckman@pha.jhu.edu,
    choopes@pha.jhu.edu, dks@pha.jhu.edu, meurer@pha.jhu.edu}
\altaffiltext{2}{Space Telescope Science Institute, 3700 San Martin Drive,
   Baltimore, MD, 21218; aloisi@stsci.edu}

\begin{abstract}

We have analyzed \fu, \xmm, and \ch ~observations of \vv, a local galaxy merger
with strong similarities to typical high-redshift Lyman Break Galaxies (LBGs). 
Diffuse thermal X-ray emission
encompassing \vv~has been observed by \ch and \xmm.  This region of hot
($\rm{kT}\sim 0.59~\rm{keV}$) gas has an enhanced
\afe~element ratio relative to solar abundances and follows the
same relations as typical starbursts between its properties (luminosity, size, and temperature) and those of the starburst galaxy
(star formation rate, dust temperature, galaxy mass). 
These results are
consistent with the X-ray gas having been produced by shocks driven by a galactic superwind.
The FUSE 
observations of \vv~show strong, broad interstellar absorption lines with a pronounced blueshifted component
(similar to what is seen in LBGs). This implies 
an outflow of material moving at $\sim300-400~\ks$ relative to \vv.  The properties 
of the strong \ovi~absorption line are consistent with radiative
cooling at the interface between the hot outrushing gas seen in X-rays and
the cooler material seen in the other outflowing ions in the FUSE data.
We show that the wind in \vv~has {\it not} created a ``tunnel'' that enables more than a 
small fraction ($<$ few percent) of the
ionizing photons from \vv ~to escape into the IGM.
Taken together, these data provide a more complete physical basis for understanding 
the outflows that seem to be generic in LBGs. 
This will lead to improved insight into the role that such outflows play in the evolution of 
galaxies and the inter-galactic medium.

\end{abstract}


\section{Introduction}

In recent years, progress in cosmology has been driven by the
remarkable results of experiments such as {\it WMAP} \citep{benn03}.  These experiments
have made precise measurements of the geometry, 
the age, and the spectrum of density fluctuations of our universe. Using
$\Lambda$ Cold Dark Matter ($\Lambda$CDM) numerical simulations, we can 
track the evolution of the large scale structure of the dark matter from the era of quantum fluctuations to
our current epoch \citep[e.g.][]{davis85,spring05}.  However, significant problems in the
simulations emerge when examining
small scale structure and regions of high density where the complex range of baryonic physics becomes important \citep{klyp99,rob05,som99}.   
The next step forward in cosmology will therefore require a much better
understanding of the physical processes involved in 
the interactions within the gas/star/black-hole ecosystem.

Feedback, powered by star formation, could play a crucial role in 
understanding galaxy formation and evolution.  This is particularly true 
in past epochs, when the overall cosmic star formation rate was significantly higher \citep{bunk04}.
In regions of intense star formation, short lived massive stars produce stellar winds and 
supernovae whose kinetic energy can collectively drive galactic scale outflows of metal enriched 
material into the galaxy halo and potentially into the Intergalactic Medium \citep[IGM,][]{aguir05}. 
These "superwinds"  can also enhance
the fraction of ionizing radiation that escapes from the galaxy into the IGM \citep{dov00}.
Therefore these complicated multi-phase winds are an important influence 
on the chemical and thermodynamical nature of the IGM. They are also likely to play a crucial role in the evolution of galaxies, particularly for low mass systems whose relatively shallow potential wells make them especially vulnerable to wind-driven loss of gas and metals \citep{trem04}.

Multiwavelength observations are required to understand the complex multiphase
nature of galactic winds. The coronal 
gas ($T \sim 10^5$ to $10^6$ K) and hot X-ray gas ($T \sim 10^6$ to $10^7$ K) in galactic winds are particularly important.  These phases
are intimately connected to the mechanical/thermal energy that drive the outflows and to the metals the outflows carry.
Far-ultraviolet (FUV) observations 
of the coronal gas provide important insights into the cooling 
and kinematics of the galactic winds \citep{heck01}.  X-ray observations 
probe the slowly cooling hot gas which is most likely to escape the
galaxy \citep{strick00}.

The most widely studied population of high redshift star forming galaxies are 
the Lyman Break Galaxies \citep[LBGs,][]{steid99}.  These galaxies can be efficiently detected for 
$2\lesssim z \lesssim 6$ using
the Lyman-break technique which picks out spectral band dropouts caused by the
912 $\ang$ Lyman continuum discontinuity. They constitute a significant (and possibly dominant) fraction of the population of star forming galaxies during this important cosmic epoch \citep{peac00}.  Galactic winds appear to be a
ubiquitous property of LBGs \citep{shap03}.  The LBGs trace the most overdense regions of the universe, which are believed to be the progenitors of present day galaxy clusters \citep{giav02}. They therefore play an important role in understanding the
evolution of such clusters.  In particular, winds from LBGs could be the process that heats
and chemically enriches the Intercluster Medium \citep[ICM, ][]{hels00,tam04}.

Unfortunately, our knowledge about the properties of the winds in LBGs is limited to what can be inferred from the interstellar UV absorption lines longward of Lyman $\alpha$. Because of their great distance, X-ray detections are almost impossible using
current X-ray observatories.  Stacking techniques have been used to
construct luminosity weighted average X-ray spectra \citep{nan02,lehm05}.  
A more fundamental problem is that the large redshifts of the LBGs mean that soft X-ray observations with {\it Chandra} or \xmm ~are observing at rest-frame energies above a few keV. In local star-forming galaxies the emission in this band is dominated by the population of X-ray binaries, and the thermal emission from the galactic wind is negligible \citep{colb04}. Likewise, observations of the coronal phase gas is very difficult in LBGs. The most accessible probe of such gas is the FUV \ion{O}{6}~$\lambda\lambda$1032,1038 doublet which lies deep within the Lyman $\alpha$ forest in the spectra of LBGs. 

It is clear that directly studying gas hotter than $10^5\,$K in high-redshift LBGs will continue to be very difficult or even impossible for the forseeable future.  Therefore, an important step in understanding galactic winds in LBGs would be to identify the best local analogs to LBGs and to then investigate their winds using the full suite of observations that are possible at low redshift.

Recently \citet{heck05} created a catalog of low redshift galaxies 
using {\it GALEX} \citep{mart05} UV observations of galaxies with spectra taken by the Sloan Digital Sky Survey \citep[SDSS, ][]{york00}. This catalog has now been significantly expanded based on additional {\it GALEX} data by Hoopes et al. (2006). They used this matched catalog to select a sample of extraordinarily rare local (z $<$ 0.3) galaxies having the same UV luminosities, sizes, and surface brightnesses as typical LBGs. They then showed that these galaxies have the same star formation rates (SFRs), galaxy masses, velocity dispersions, and chemical compositions as typical LBGs. Based on the sample observed to date with {\it GALEX}, \vv~  at z $\sim$ 0.02 is the 
closest known LBG analog, thus making it ideal for X-ray and FUV studies.

{\it HST} NICMOS and STIS observations by \citet{gold02} of \vv~(ARP 236) have
shown two distinct interacting/merging components  to the galaxy.  The two 
components are separated by 6 kpc in projection ($\sim 15 \arcsec$).  The eastern 
component, while prominent in the NIR NICMOS images, is almost invisible in the STIS UV images
due to heavy dust obscuration. The western component is very bright in the UV, 
with a far-UV luminosity of $2.2\times10^{10}\,\rm{L_\odot}$, half light radius of 2.3 kpc, and 
effective mean surface 
brightness of $6\times10^8~{\rm L_\odot/kpc^2}$, all comparable in value to those of 
LBGs \citep{heck05}.  CO (1-2) observations by \citet{yun94} show that the molecular gas is 
centered between
the two components of \vv.  Their measured velocities of the molecular gas
coincident with the western component 
show a rotating disk with a velocity range between $5900~\ks$ and $6200~\ks$.
\citet{yun94} also show, from 1.4 GHz radio continuum observations, that
star formation is spread throughout the \vv ~system.   The NIR, MIR and CO (3-2) emission of \vv~
is dominated by a source at the center of the eastern component.  
Combinations of compact star formation \citep{iono04} and an obscured AGN \citep{floc02} have been used to explain the emission.  The observed  IRAS $\rm{L_{FIR}\sim2.8\times10^{11}\, L_\odot}$ 
\citep{soif89} is then a measure of the entire system's SFR and implies a total SFR of 
$\sim48\,\rm{M}_\odot$ per year \citep{kenn98}.  \citet{gold02} suggest that the star formation
is split roughly equally between the two components.  These properties are summarized in Table \ref{prop}.

\section{The Data}

\subsection{\fu~Observations \& Analysis\label{fusetalk}}

The \fu~spacecraft is composed of four separate mirrors and two detectors \citep{moos00}.  Two mirrors
are coated with LiF and two with SiC.  The LiF 
mirrors have better sensitivity in the $1000~\ang \lesssim \lambda  \lesssim 1180~\ang$
wavelength region while the SiC are optimized for observations in the range
between $900~\ang$ to $1000~\ang$.  Observation pointing is controlled
by a Fine Error Sensor (FES) camera located along the LiF1 channel
which results in a pointing rms of $\sim0.5 \arcsec$.

\begin{figure}
\centering
\leavevmode
\includegraphics[width=4in]{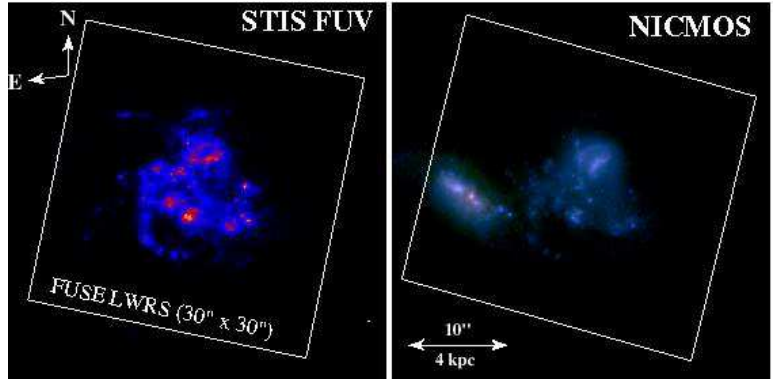}
\caption{
FUSE LWRS aperture overlayed on a HST STIS FUV and NICMOS falsecolor  image.  
Only the western component is visible in the FUV image.  Both images are from \citet{gold02}
and use a logarithmic scale map.  The NICMOS falsecolor image is an overlay of the
F110W (1.025\micron), F160W (1.55\micron), and F222M (2.3\micron) images as the 
blue, green, and red respectively.
\label{stis}}
\end{figure}

\fu observed \vv~($\alpha_{2000}=01^h07^m46^s.60$,
$\delta_{2000}=-17^\circ30^m24^s.0$) using the LWRS 
aperture ($30\arcsec \times 30\arcsec$) on July 26, 2003.  The 
observations were centered on the FUV emitting western component
of \vv.  The \fu aperture is shown overlayed on the {\it HST STIS} FUV (F25SRF2)
observation of \vv~in Figure \ref{stis}.

The \fu observation of \vv~was split into five exposures.  For every exposure and detector segment
a raw time-tagged event list was produced.  We processed each event list using the
latest \fu calibration software, CALFUSE v3.1.2 \citep{dix03}.  This software applies flux and wavelength
calibrations while correcting for a variety of effects including instrument motion, geometric distortions, spacecraft doppler shifts, event bursts, and the South Atlantic Anomaly (SAA).  

\begin{figure}
\centering
\leavevmode
\includegraphics[width=3in,angle=90]{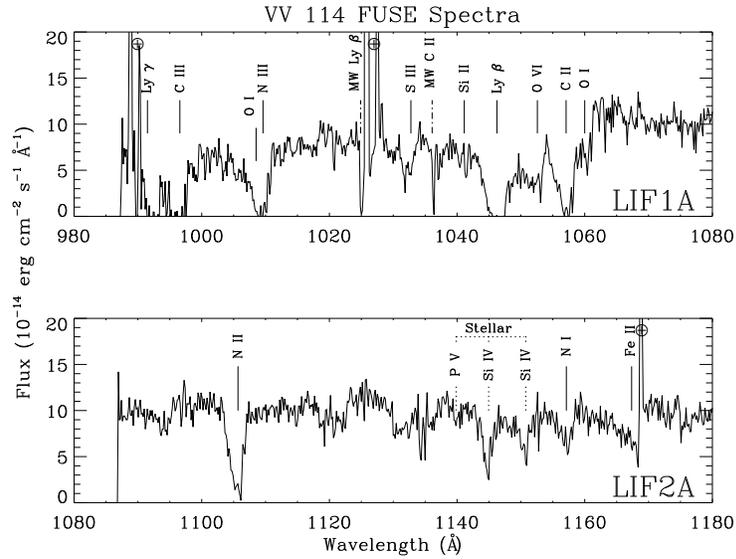}
\caption{
Smoothed spectra of \vv~from the LiF1A and LiF2A channel segments of \vv.  Prominent ISM, stellar
photospheric, milky way, and airglow lines have been marked.
\label{2spec}}
\end{figure}

\begin{figure}
\centering
\leavevmode
\includegraphics[width=3in,angle=90]{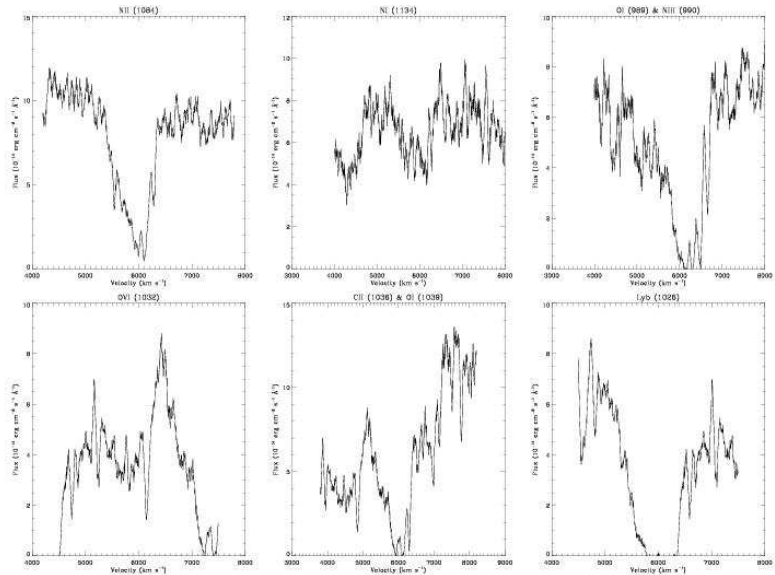}
\caption{
Absorption line profile of some of the most prominent lines.   Two blended absorption lines have
been plotted,  \ion{C}{2}~$\lambda$1036 and \ion{O}{1}~$\lambda$1039, and \ion{O}{1}~$\lambda$989 and \ion{N}{3}~$\lambda$990.  The spectra have been smoothed by a factor of 8.
\label{multipleabs}}
\end{figure}

The Hydrogen Lyman series absorption lines dominate the spectrum 
below $\sim1000~\ang$ so we have
focused on the longer \fu~wavelenths.  As the LiF channels have a greater
sensitivity and thus better S/N we have ignored the SiC channels in
our data analysis.  A comparison of the LiF1 guiding channel to the LiF2
channel confirms that channel drift is negligible for our data.
In order to co-add the spectra we cross correlated the spectra to correct for
small wavelength shifts.  Also, as the count rate is significantly lower 
in the first exposure, we have excluded it from the final combined
spectra for each channel.  This produces four combined spectra, LiF1A and LiF2B covering
$\sim985~\ang-1075~\ang$ and LiF1B and LiF2A in the range between $\sim1090~\ang-1185~\ang$.
The best spectrum in each of the two wavelength regions is shown in Figure \ref{2spec}.  
We have also labeled the most prominent FUV ISM absorption lines, stellar photospheric lines, Milky Way absorption lines, and airglow features.  The spectra around several of the most prominent ISM absorption lines are shown in Figure \ref{multipleabs}.

\begin{figure}
\centering
\leavevmode
\includegraphics[width=3in]{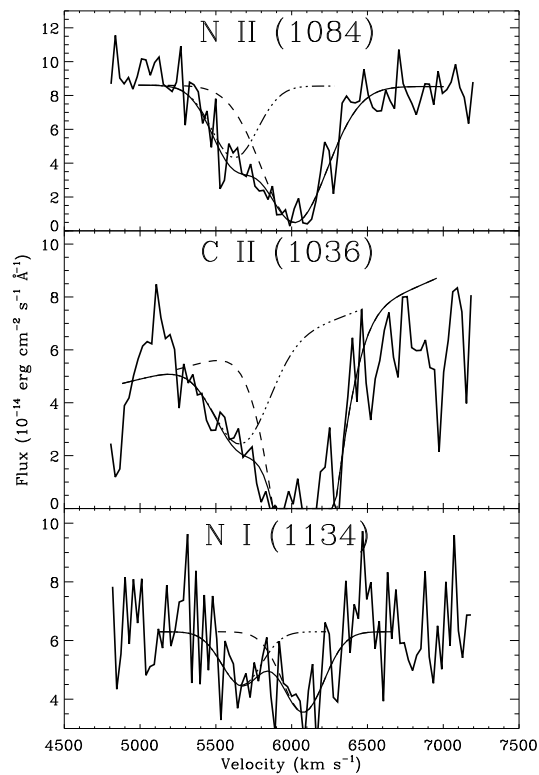}
\caption{
These three absorption lines show several similar 
absorption line features.  All three lines have a blueshifted component
in addition to a stronger component associated with the galaxy.  The dashed lines represent
the two absorption lines fit to the data while the solid line is the total fit.  Note that the \ion{C}{2}~$\lambda$1036
is mildly blended with the \ion{O}{1}~$\lambda$1039. For comparison, the
galaxy systemic velocity based on mm-wave CO
observations is $v_{sys}\simeq6050~\ks$ while stellar photospheric lines
in the FUSE spectrum yield $v_{sys}\simeq5970~\ks$.  The spectra have been binned by a factor of 10.
\label{3abs}}
\end{figure}

An examination of the spectrum of \vv~shows that it is dominated by
broad ISM absorption lines.  The strongest lines (e.g. Ly~$\beta$, \ion{C}{3}~$\lambda$977, \ion{C}{2}~$\lambda$1036) are saturated in their cores.
Most of the strong, higher S/N lines show common structures in their profiles.
This can be seen in Figure \ref{3abs} where we have vertically aligned the \ion{N}{2}~$\lambda$1084, \ion{C}{2}~$\lambda$1036,
and \ion{N}{1}~$\lambda$1134 absorption lines.  Some of the small scale
absorption features are found at $\sim$6300, 6100, and 5900 $\ks$.  The $6300\,\ks$
component is interesting as it could be infalling material or 
non-circular motion from the ongoing merger of the two galaxy components.
More important however is the general shape of the absorption line profiles.  All three lines in Figure \ref{3abs}
are asymmetric, with a pronounced wing of blueshifted absorption extending 
down to $\sim$5000 $\ks$.

To fit the lines we used the iraf tool {\it specfit} \citep{kriss94}.  Initially, each line was fit
separately using a freely variable powerlaw for the continuum and a symmetric gaussian absorption line.
As we have
double coverage of most wavelength regions, we 
independently measure most of the absorption lines and continuum twice.
For many of the lines, the line width (FWHM) is not well
constrained so we have fixed it to $700~\ks$
which is a value consistent with our fits to
the strong lines.  Some additional complications
are also important to note.  The \ion{C}{2}~$\lambda$1036 and \ion{O}{1}~$\lambda$1039
lines are mildly blended so we have fixed the 
central velocity of the \ion{O}{1}~$\lambda$1039 line relative to that of the
\ion{C}{2}.  We then tied the two FWHMs together
and fit the two lines.  
We used the same approach with the two strongly blended lines,
\ion{O}{1}~$\lambda$989 and \ion{N}{3}~$\lambda$990.  Although we were able
to fit these two lines, they are badly blended and so there is a
significant uncertainty in our derived parameters.  Table \ref{absdata} shows
the results of our fits to the data.  Errors are one $\sigma$ and are
calculated from a rescaled inversion of the curvature matrix.  
There are a few weak lines
in the spectra that we have not fit due to
their low S/N.

The single component fits above do not account for the blueshifted component clearly seen in the 
higher S/N absorption lines.  Therefore, for the strongest lines 
(e.g. \ion{C}{2}, \ion{N}{2}, \ion{N}{1}, \ion{O}{1})
we have refit the data with a second blueshifted gaussian absorption line.  
We have excluded the strongly blended lines \ion{O}{1}~$\lambda$989 and \ion{N}{3}~$\lambda$990 and the heavily saturated Lyman$\beta$ and \ion{C}{3}~$\lambda$977 lines from
these new fits.  \ion{O}{6} was also excluded as it clearly has a different
line profile than the other strong lines (see below).
The 2-component fits
are shown in Figure \ref{3abs} and Table \ref{twoabsdata} and significantly improve the fit quality of each line.  The results for the different lines 
are all consistent with a strong absorption
line at  $\sim6050~\ks$ with a second blueshifted line at  $\sim5650~\ks$.  The blueshifted component
generally has an equivalent width of typically about 0.4 that of the principal component and a marginally
smaller FWHM.
The success of the two absorption component fits suggests an inadequacy
in the single line fits.  This provides a caution in using the single line fits to determine 
central velocities of the absorption lines.  The blueshifted component is probably
a general feature, but is not conspicuous in the weaker lines due to their lower S/N.

%Two prominent stellar photospheric \ion{Si}{4}~(1122 \& 1128) absorption 
Several prominent stellar photospheric absorption
lines are observed in the data.  These lines allow us to independently constrain the
velocity of the UV emitting portion of \vv.  We have fit these
absorption lines in the same manner as described above.  The line
centroids were tied together in order to strongly constrain
the relative velocity.  Our fit results can be found in Table \ref{phabsdata}.
The fitted FWHMs of the photospheric lines are significantly narrower at
$390~\ks$ than the ISM FWHMs ($\sim700\,\ks$).
Combining the measurements we find a system
velocity of $5971\pm14~\ks$.

\subsection{\ch~Observations\label{chandraobs}}

\vv~was observed by \ch on October 20, 2005.  The data was obtained using very faint
mode with the galaxy centered on the ACIS S3 aim point.  Standard data
processing was applied using CALDB 3.1.0 and ASCDSVER 7.6.3 \citep{ciao}.
Less than 1 ks of observation time was lost due to
anomolous background levels leaving 58.8 ks of total integration time. 

For image and spectral analysis we followed the same procedures as 
previously described in \citet{grim05}.  Using an image of the counts in the 
0.3 - 1.0 keV range we calculated a
90\% counts enclosed radius for \vv~of 4.4 kpc.  The 0.3 - 1.0 keV band was chosen as it 
is sensitive to the diffuse thermal emission and minimizes point source contamination.
We also created 
adaptively smoothed images using the {\rm CIAO} task {\rm CSMOOTH}.
For all of our spectral analysis a local background was used.  The background was
centered on the galaxy but excluded all diffuse and point source emission.  Spectra
were extracted using the {\rm CIAO} script {\rm SPECEXTRACT} and grouped 
using a minimum of 20 counts per bin.

\begin{figure}
\centering
\includegraphics[width=4in]{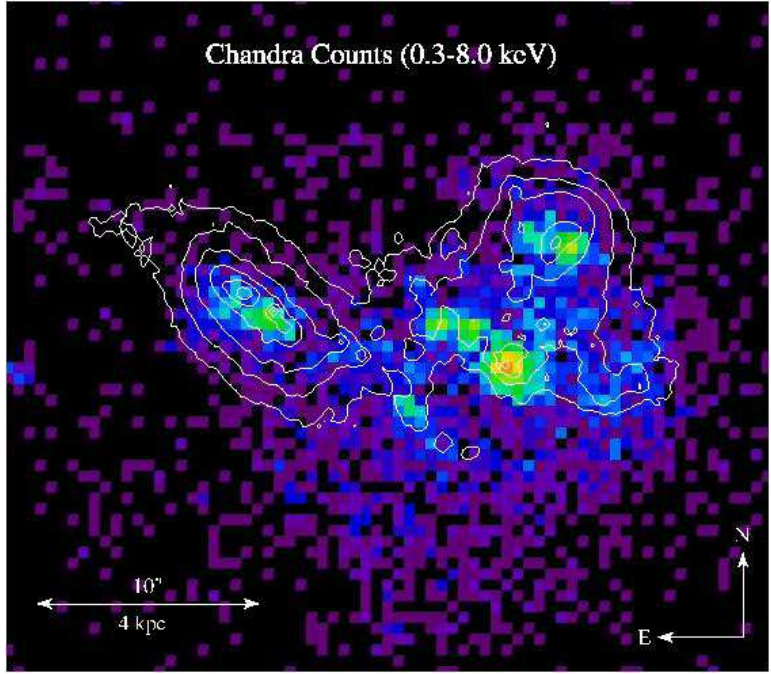}
\caption{
Here we have displayed the \ch~0.3-8.0 keV X-ray counts image.  The NICMOS F160W (1.55\micron)\,image of \vv~is overlayed for comparison.  Both are displayed using a logarithmic
scale.  The western component of \vv~dominates the X-ray emission.  The
X-ray emission of the western component has a similar morphology to the FUV STIS image in
Figure \ref{stis}.  Diffuse X-ray emission 
can also be seen extending south of the galaxy.  The eastern component of the galaxy
is also clearly seen in the X-ray image.
\label{chandracnts}}
\end{figure}

The high spatial resolution \ch~counts (0.3-8.0 keV) image in Figure \ref{chandracnts} has 
several interesting features.
Both the eastern and western components are clearly observed.  
The western component dominates the X-ray emission and has a similar morphology to
that seen in the STIS data. The peak of the X-ray emission coincides with the brightest star cluster seen
in the UV.  Diffuse X-ray emission is seen to the south of the galaxy and extends far beyond 
the limits of the FUV and NIR emission.  

\begin{figure}
\centering
\includegraphics[width=4in]{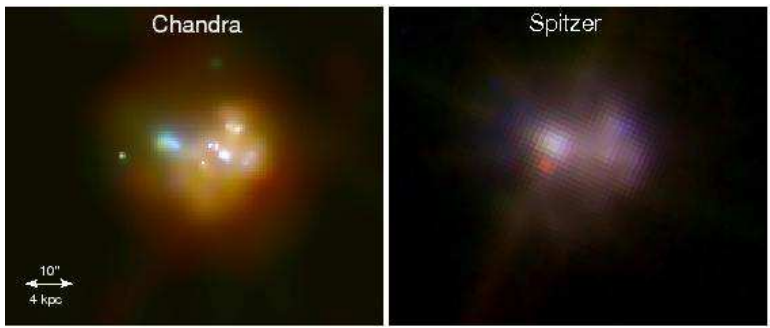}
\caption{
\ch and Spitzer IRAC false color images of \vv.  The \ch image has been adaptively 
smoothed and constructed using 0.3-1.0, 1.0-2.0, and 2.0-8.0 keV as the red, green, and blue
colors respectively.  Red, green, and blue are mapped to channels 4 (8\micron), 3 (5.8\micron), 
and 1 (3.6\micron) in the Spitzer IRAC image.  The Chandra image shows strong absorption
of the emission from the eastern component.  An point source is coincident with the
unresolved source also seen at the center of \vv E in the NICMOS and Spitzer images. 
\label{chandrafalse}}
\end{figure}

\begin{figure}
\centering
\includegraphics[width=3in]{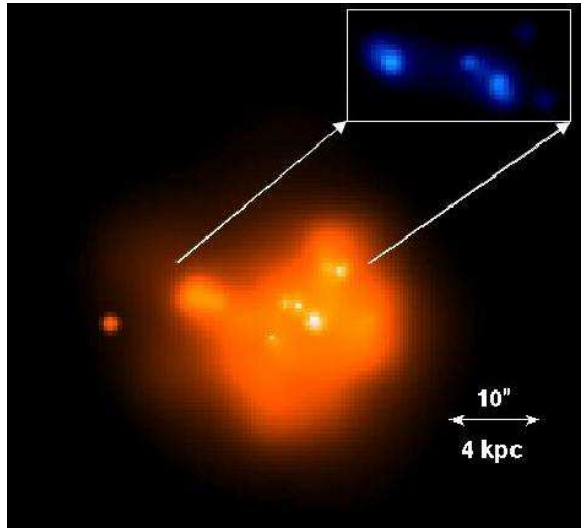}
\caption{
A smoothed soft band (0.3-2.0 keV) \ch~image of \vv~shows that soft emission
extends far beyond the central regions.  A blown up image
of the 3-8 keV central region is displayed in the upper right corner.  
This small region contains all of the observed hard X-rayemission in \vv.  In the hard X-ray
image \vv E has a higher peak brightness and is significantly more compact
than \vv W.  They have comparable hard X-ray fluxes while the heavily absorbed 
\vv E is barely observed in the soft band.  
Both images are
plotted on a log scale.
\label{all_vs_east}}
\end{figure}

Figure \ref{chandrafalse} shows an adaptively smoothed Chandra false color image of the
galaxy.  The emission from \vv E is harder than that from the western component.  
This is consistent with the high column densities suggested by the FUV observations 
(Figure \ref{stis}).  Soft X-rays from \vv E are absorbed so that, as in the FUV, the soft
X-ray emission is primarily tracing \vv W.   Figure \ref{all_vs_east} has a soft band
image of the entire galaxy with a blown up picture of the inner regions
in the hard band.   A compact source is also  
detected near the center of \vv E.
It is coincident with the unresolved
source seen in the NICMOS and lower resolution Spitzer IRAC images.

\begin{figure}
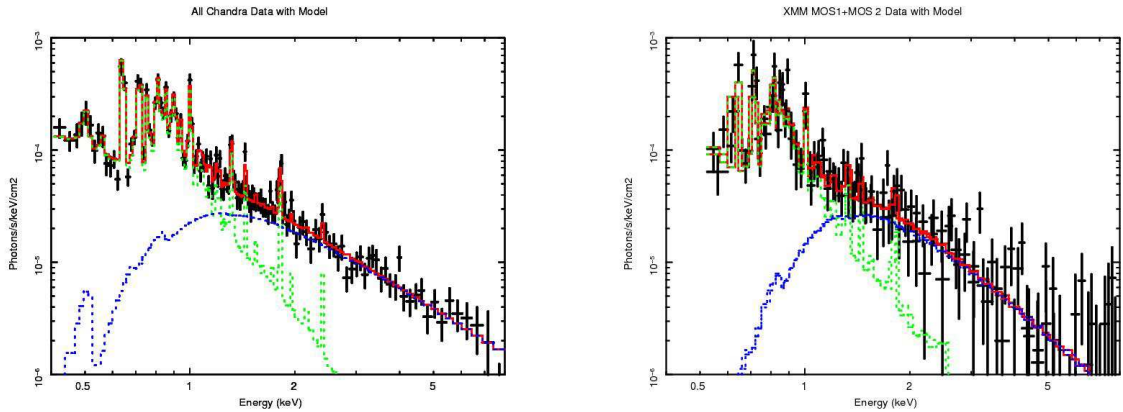

\centering
\columnwidth=.4\columnwidth
\includegraphics[width=2.4in,angle=-90]{f8a.jps}
\hfil
\includegraphics[width=2.4in,angle=-90]{f8b.jps}
\vspace{.01in}
\caption{
The total \ch and \xmm~MOS+MOS2  unfolded spectra of \vv~with total fit (red), thermal vmekal component (green), and powerlaw component (blue).  The spectra are comparable and are both well modelled by a thermal plasma ($\rm{kT\sim0.6\,keV}$) plus a powerlaw with intrinsic absorption.
\label{xspec}}
\end{figure}

The spectra were well fit by a model that included an absorption column, 
a thermal plasma model, and an absorbed powerlaw.  XSPEC's
vmekal model was used which allows for variable metallicities. 
The derived model fit parameters are listed in Table \ref{xspecdata}
while the data and fit are plotted in Figure \ref{xspec}.
Following \citet{grim05} we have only calculated the 
\afe~ratio due to degeneracies in the model fit in determining the absolute elemental abundances.
The \afe~ratio of $2.7^{+0.6}_{-0.7}$ (90\% confidence)
assumes the default XSPEC solar abundances of \citet{and89}.  More
recent abundance measurements from \citet{asp05} obtain comparable results
with an \afe~ratio of  $2.5^{+0.8}_{-0.3}$ .  
The 0.3 - 2.0 keV luminosity of the
thermal emission  \vv E is $2.2\times10^{41}$ \ergs.  

\subsection{\xmm~Observations}

The \xmm~data was retrieved from the US public archive
at HEASARC, and analyzed using version 5.3 of the Standard
Analysis System (SAS). The latest calibration files available at the
time were used. Images and spectra were extracted from the MOS1 and MOS2
events files produced by the XMM-Newton data pipeline, with total
exposures of 11367 and 11348 seconds respectively. No data from the PN
detector was available.
The spectra of VV 114 included all events within a radius
of $30\arcsec$ of the peak in the soft X-ray emission seen
in the X-ray images, which closely coincides with the NED
position for VV 114. A background spectrum was taken in an annulus of
inner and outer radii $40\arcsec$ and $80\arcsec$
respectively. The spectra was subsequently analyzed using XSPEC 11.2. 

We simultaneously fit the MOS1 and MOS2 spectra using the same model
as for the \ch~data.  
Table \ref{xspecdata} lists the derived model parameters while
Figure \ref{xspec} shows
the unfolded spectra and model components.  
%The combined MOS1+MOS2 spectra contains $\sim1500$ counts while
%the longer \ch~observation has over 5000 counts.  
The results of
the \xmm~fit are consistent with those from the longer exposure, higher quality \ch~data.
 The derived vmekal plasma temperature is $0.59^{+0.06}_{-0.09}$ keV with an
\afe~ratio of $2.2^{+2.0}_{-0.7}$ (abundances from \citet{and89}).  
The  \citet{asp05} abundances obtain similar results
with an \afe~ratio of  $1.5^{+1.9}_{-0.6}$ (90\% confidence).
We've also derived the absorption corrected luminosity
of the thermal plasma as $2.0\times10^{41}$ \ergs ~in the 0.3-2.0 keV energy
range.

\section{Discussion}

\subsection{AGN and Starburst Activity in \vv E\label{vv114star}}

While the western component of \vv~dominates
the UV and X-ray emission, \vv E is the more powerful
near-IR emitter (right panel in Figure \ref{chandrafalse}).  
NIR spectral observations by  \citet{doy95} show
evidence of a gas photo-ionized by 
$\sim4\times10^5$ OB stars in \vv E.  
The extended emission
observed in the radio \citep{con91} and
NIR \citep{doy95} bolster the argument 
that a compact starburst dominates \vv E.
\citet{floc02} however observed a continuum bump at 5-6 \micron\,
which is typical of AGN spectra. They suggest that a heavily absorbed AGN
could contribute up to 40\% of the mid-IR flux in \vv E.

\begin{figure}
\centering
\includegraphics[width=3in,angle=-90]{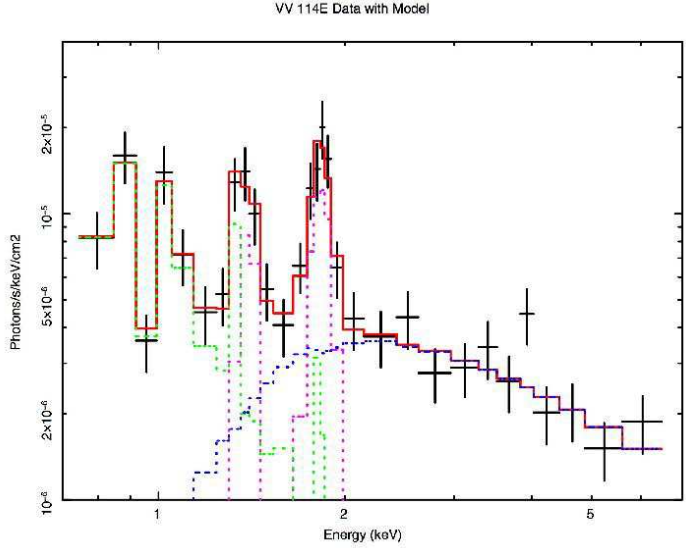}
\caption{
The \ch ~unfolded spectra of \vv E with total fit (red), thermal vmekal component (green), powerlaw component (blue), and two gaussians (purple).  The powerlaw is fairly flat with a photon index of 1.3 and intrinsic absorber with $\rm{N_H\sim2\times10^{22}\,cm^{-2}}$.  The two gaussians fit  the enhanced Si ($\sim1.4$\,keV) and Mg ($\sim1.8$\,keV) which are typical in photo-ionized regions.
\label{vv114e}}
\end{figure}

We extracted \ch~spectra of a small circular region surrounding
\vv E.  The extracted spectra can be seen in Figure \ref{vv114e}.  This
region encloses roughly 600 counts, about 10\% of the total
counts observed in \vv.  The spectra is heavily absorbed requiring
 $\rm{N_H\approx2\times10^{22}\,cm^{-2}}$.  A fairly flat powerlaw is also fit
with a photon index of $\sim1.3$, consistent with the existence of a buried AGN.
%However the binned spectra show no sign of
%a strong Fe~K$\alpha$ line, which would provide some evidence of an AGN.

We initially fit the lower energy spectra of \vv E with a simple absorbed 
thermal plasma model with variable abundances.  This 
provides a poor fit and requires abundances of over 13 times solar values.
The super-solar abundances are driven by the enhanced Si ($\sim1.4$\,keV) and Mg ($\sim1.8$\,keV)
line emission.  This is typical in photo-ionization regions
and is seen in stellar winds from high-mass X-ray binaries \citep[HMXBs,][]{sak02},  OB stars \citep{osk06},
and some obscured AGN \citep{lev06}.  
The addition of two gaussians to the model fit
 results in reasonable sub-solar abundances and a significantly
better fit with a reduced $\chi^2\sim0.8$ (Figure \ref{vv114e}).  

The thermal plasma dominates the emission in the 0.3-2.0 keV range with 
an unabsorbed luminosity of $2\times10^{41}\rm{\,ergs\,s^{-1}}$.  
The eastern component's thermal plasma luminosity of $2\times10^{41}\rm{\,ergs\,s^{-1}}$  
is comparable to
the luminosity we inferred for the {\bf entire} galaxy in section \ref{chandraobs}.
As the observed photons are primarily from the unabsorbed western component
a successful model can ignore the heavily absorbed eastern component.
However, this causes us to underestimate the soft 
X-ray luminosity of \vv~by a factor of $\sim2$. 
 This complication does not occur in the
harder X-ray range where we derive the 2-10 keV luminosities for the eastern and 
western components as $1.1\times10^{41}\,\rm{erg\,s^{-1}}$ and 
$1.3\times10^{41}\,\rm{erg\,s^{-1}}$ respectively.  
Previous works \citep{bauer02,franc03} have shown that the
hard X-ray emission can be used as a SFR tracer.  
Assuming the hard X-ray emission is tracing HMXBs and not contaminated by an AGN  
 we can derive the SFR for the two 
galaxy components.  This implies that they have 
a roughly equal star formation rate 
with $\sim28\,\rm{M_\odot/year}$ for \vv E and  $\sim33\,\rm{M_\odot/year}$ for \vv W \citep{franc03}.  
The total star formation rate of $\sim61\,\rm{M_\odot/year}$ (30\% scatter) is 
consistent with the $\sim48\,\rm{M_\odot/year}$ inferred from the FIR luminosity
suggesting that star formation is the dominant energy source.

An AGN contribution however is still not ruled out.  Although $\rm{Fe\,K\alpha}$ emission
is not observed in the binned (minimum 20 counts) spectra of \vv E, we have extracted
an image in the 6.25-6.55 keV range.  A single feature, centered on the unresolved
source in the \ch~and NICMOS images, is visible.  The five detected counts are well above the
expected contribution from the background plus the fitted powerlaw of less than 0.01 counts.
We extracted unbinned spectra from circular region with a diameter of 2.3\arcsec (1 kpc) and centered
on the unresolved source in \vv E.  We fit the spectra with a gaussian centroided at 6.4 keV and 
found a best fit equivalent width
of 0.3 keV.  As the gaussian is poorly constrained, we have instead found a 90 \%
confidence level upper limit on the equivalent width of 1.3 keV.  The 
luminosity in the $\rm{Fe\,K\alpha}$ line is $1.3\times10^{40}\,\rm{erg\,s^{-1}}$. 

\citet{lev06} have studied a sample of heavily absorbed AGN and found that the IR and
$\rm{Fe\,K\alpha}$ emission line luminosities are correlated with
 $\log (\rm{L_{IR}/L_{Fe\,K\alpha})\approx 3.5}$.  For \vv~this would suggest a 90 \% confidence upper
 limit to the AGN contribution of $\rm{L_{IR}}<1\times10^{10}\,\rm{L_\odot}$.  This is significantly less than the 
 $\rm{L_{IR}}\sim5\times10^{11}\,\rm{L_\odot}$ observed by IRAS.  
 This supports the idea that a compact nuclear starburst dominates
 the central region of \vv E.
 The weak detection of $\rm{Fe\,K\alpha}$ is also consistent with the nuclear starburst argument.  
 We note that $\rm{Fe\,K\alpha}$ lines are detected in HMXBs \citep{sak02}.  Assuming
 that the 2.0 - 10 keV luminosity is not contaminated by an AGN, almost one quarter of the SF in the entire
 galaxy is concentrated within a region 1 kpc in diameter at the center of \vv E.

We conclude that while a highly obscured AGN {\it may} be present in the nucleus of \vv E, it would not
contribute significantly to the bolometric luminosity of the \vv~system. The energetics of \vv~are instead
dominated by the intense episode of star formation it is undergoing.

\subsection{\fu Observations of the Wind}

The success of the two component fits to the absorption line profiles suggest a simple dynamical picture.  The
stronger absorption feature at $\sim6050~\ks$ traces the ISM of \vv W.  This velocity is consistent with that of the molecular gas of the western
component.  The stellar absorption lines, which we would also expect to be associated with 
the western component, appear to be at a slightly lower central speed of $\sim5970~\ks$.
This difference in velocities could be explained by a systematic error in the two component
absorption model caused by degeneracy in some of the fit parameters.  A difference in the relative spatial distribution between the hot stars and ISM and the way they sample the galaxy rotation curve
could also explain this small
velocity shift.
The blueshifted ISM absorption feature has a mean velocity of $\sim5650~\ks$, implying an outflow of gas moving at a mean velocity of $\sim$300-400 ~$\ks$ away from the galaxy. Weaker absorption can be seen at velocities as low as $\sim5000~\ks$ (e.g. \ion{N}{1} and \ion{O}{1}), suggesting that the wind is accelerating clouds up to outflow velocities as high as $\sim1000~\ks$.

It is important to compare our results to previous absorption line
studies of LBGs.  In the composite LBG spectrum created by 
\citet{shap03} they measure broad (FWHM$\sim550~\ks$) and
blueshifted ($v~\sim150~\ks$) ISM absorption lines.  In the observations
of CB58 by \citet{pett02} they find a mean outflow speed of $v=255~\ks$ and slightly
broader lines.  Both these results are consistent with what we observe
in \vv.  

It is also interesting that the kinematics of the coronal gas as traced by the 
\ovi ~absorption line is different from what is seen in the cooler gas.
In the single absorption component fits
the centroid of the  \ovi ~absorption line is blueshifted by about 100 $\ks$ compared to the other strong lines, and the line profile is much more symmetric.
Absorption in \ovi ~can be seen at velocities as low as $\sim5000 ~\ks$, consistent with outflow velocities as high as $\sim1000~ \ks$. Unfortunately, the blue wing of the \ovi ~line merges with the red wing of the adjacent Lyman $\beta$ line, making it impossible to tell if even faster moving coronal gas might be present.

The different kinematics of the \ovi ~line implies a different physical origin.
The \ovi ~line is a particularly interesting one
due to its importance in radiative cooling and the relatively 
narrow temperature range over which it is a significant ion. \citet{heck01}
also observed \ovi ~absorption in the dwarf starburst
NGC~1705.  In NGC~1705 the \ovi~ line was also more strongly blueshifted than the other ions.    
They attributed the production of  \ovi~
to the intermediate temperature regions created by the 
hydrodynamical interaction between hot outrushing gas
and the cool fragments of the ruptured superbubble seen in H$\alpha$ images. Such a situation is predicted to be
created as an overpressured superbubble 
accelerates and then fragments as it expands out of the galaxy.

\begin{figure}
\centering
\includegraphics[width=3in]{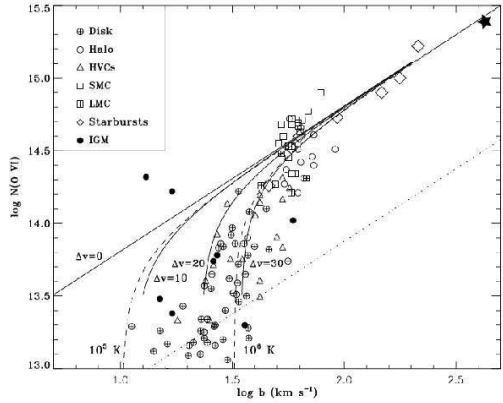}
\vspace{.2in}
\caption{
Column density vs. line width for a wide variety of \ovi~absorption line systems including
galactic disk and halo, high velocity clouds, starburst galaxies, and the IGM.
The two dashed lines indicate the predictions  for radiatively cooling gas for assumed
temperatures of $\rm{T_{OVI}=10^5~\&~10^6~K}$.  This plot is originally from \citep{heck02}.
\vv~has been added and is represented by the star in the upper right.  \vv~extends the relationship
between flow velocity and column density to significantly higher values.
\label{ovi}}
\end{figure}

\citet{heck02} derive a simple and general relationship between the \ovi~
column density and absorption line width which will hold whenever there is a radiatively cooling gas flow passing through the coronal
temperature regime.  They showed that this simple model accounted for the properties of \ovi ~ absorption line systems as diverse as clouds in the disk and halo of the
Milky Way, high velocity clouds, the Magellanic Clouds, starburst outflows, and the clouds in the IGM. Using the
measured equivalent width in Table \ref{absdata} we have measured an
\ovi~column density of $\rm{\log(N_{OVI})=15.3}$.  Figure \ref{ovi}
shows the predictions and data from \citet{heck02} and includes \vv.  
\vv~has the both the highest \ovi~ column density and broadest line width in the plot, and is fully consistent with
the model predictions. It would then trace radiatively cooling gas in a high speed outflow.

For the outflow components, it is interesting to estimate the mass and energy outflow rates. We follow 
\citet{heck00} and assume a mass-conserving
outflow with a constant velocity. This allows us to estimate the mass and
energy outflow rate in the wind from the total gas column density, outflow
velocity and starburst radius (equations 5 and 6 in \citet{heck00}).
To estimate the gas column density in the outflow we use ionic column 
densities for the blueshifted
wind component derived from the strongest unsaturated lines. We then convert
these into hydrogen column densities assuming solar abundances from 
\citet{asp05}.  The Oxygen abundances of $12 +$ log[O/H] $=$ 8.6 and 8.7 \citep{kim95,char01} for 
the two most prominent 
knots in \vv W suggests that this is reasonable.
The hydrogen column density in the
neutral gas can be measured from both \ion{N}{1} and \ion{C}{2}.
Both predict a column density of about $\sim6\times10^{18}\,\rm{cm^{-2}}$.
The \ion{N}{2} column density leads to a corresponding ionized hydrogen
column density of $1.4\times10^{19}\,\rm{cm^{-2}}$. This is a lower limit
since we make no correction for more highly ionized nitrogen.
Finally, 
\ion{O}{6} predicts a lower limit to the hydrogen column of the
coronal phase of the outflow of $2.3\times10^{19}\,\rm{cm^{-2}}$,
based on the maximum possible \ion{O}{6} ionic fraction of $\sim$20\%
for collisional ionization equilibrium \citep{suth93}. The lower bound
to the total hydrogen column density in the outflow is then 
$\sim4\times10^{19}\,\rm{cm^{-2}}$. Taken a starburst radius of 2.3 kpc
($\rm{R_{50,UV}}$), and an outflow velocity of $400\,\ks$ we find 
an mass outflow rate of $\sim5\,\rm{M_{\odot}}$ per year and a kinetic
energy outflow rate of $\sim2.5\times10^{41}$ erg/sec (assuming
a wind opening angle of 4$\pi$ steradians).
The derived values are in
Table \ref{outflow}. To put these values into context, in section \ref{vv114star}
we estimated the star formation rate of \vv W as roughly 24-33 $\rm{M_{\odot}}$.
The implied star formation
rate is of order the outflow rate. The
total rate at which supernovae and stellar winds provide mechanical
energy in VV~114W is $\sim6\times10^{42}$ erg/sec, and the estimated
energy outflow rate is therefore only $\sim$5\% of the total. These results
are similar to what is seen in typical far-IR luminous starburst galaxies
based on the interstellar NaI D absorption-line \citep{heck00,rup05,mart05b}.

\subsection{Physical Properties of the Hot Phase of the Wind}

We can use the parameters for the fit to the total Chandra and XMM
X-ray spectra of the diffuse gas in \vv~
to derive estimates for the basic physical properties of the hot phase
of the wind. The normalization for the VMEKAL
component implies an emission integral (the volume integral of density
squared) of $2.4 \times 10^{64}$ cm$^{-3}$. For the geometrical volume
of the emitting region we take a sphere whose radius encompasses
90\% of the soft X-ray emission ($\sim$4.5 kpc). This then implies
a mean gas density of $n \sim 5\times 10^{-2} f^{-1/2}$ cm$^{-3}$ and
a gas mass of $M \sim 4 \times 10^{8} f^{1/2}$ M$_{\odot}$
(where $f$ is the volume filling factor of the X-ray emitting material).
For kT = 0.59 keV
the mean thermal pressure is $P = 1.3 \times 10^{-10} f^{-1/2}$ dynes cm$^{-2}$
and the total thermal energy content of the hot gas is $E = 1.4 \times 10^{57}
f^{1/2}$ ergs.

Taking the characteristic timescale to be the above radius (4.5 kpc) divided by the
sound speed in the hot gas ($\sim$ 500 km/sec) yields an age of 9 Myr. The implied
outflow rates in the hot gas are then $\dot{M} \sim 50 f^{1/2}$ M$_{\odot}$/year and
$\dot{E} \sim 5 \times 10^{42} f^{1/2}$ erg/sec. For volume filling factors similar
to those estimated for other starburst winds ($f \sim$ 0.1 to 1),
the implied mass outflow rate in \vv~is comparable
to the total star formation rate ($\sim$ 30 M$_{\odot}$/year) and the rate of energy transport is 
comparable to the total
rate at which mechanical energy would be supplied by supernovae and stellar winds
in the starburst ($\sim 10^{43} f^{1/2}$ erg/sec). These outflow rates significantly exceed
the rates for the cooler gas derived from the $FUSE$ data above, but are
are typical of winds in powerful local
starburst galaxies \citep[e.g.][]{heck03}.

\subsection{X-ray Properties Compared to SF Galaxies}

In \citet{grim05} we studied the X-ray properties of 22 star forming
galaxies spanning over 4 orders of magnitude in X-ray luminosity 
including \ulg ~like Arp~220, normal starbursts like M~82, and dwarf starbursts like NGC~1705.  Our principal
conclusion was that the properties of the 
hot gas in all of these star forming galaxies were remarkably similar.  
In general, simple
scaling relations were able to explain many of the differences between the
galaxies in the sample.  These relationships are consistent
with the basic superwind scenario in which the mechanical energy deposited by stellar winds and supernovae
results in
an over-pressurized cavity of hot gas within the
starburst.  The cavity will expand, driving the hot gas
outwards, possibly out of the galaxy and into the surrounding IGM.   

\begin{figure}
\centering
\includegraphics[width=2in,angle=-90]{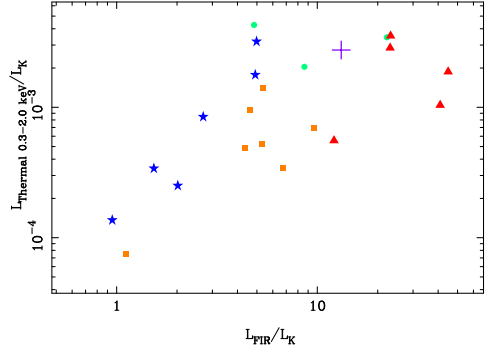}
\vspace{.2in}
\caption{
In this plot taken from \citet{grim05} we have divided both the thermal X-ray (0.3-2.0 keV)
and the FIR luminosity by the K-Band luminosity (a proxy for stellar mass).  The FIR luminosity
of the dwarf starbursts is actually the sum of the their UV and FIR luminosities.
Their is a clear
linear relation between the SFR per stellar mass and
the thermal X-ray emission per stellar mass.  The IRAS and K-Band
data was obtained from {\it NED} and have been tranformed using
the methods of \citet{sand96} and \citet{carp01} respectively.  \vv ~ follows the scaling
relation defined by starbursts.
Key: Orange Squares-Dwarf Starbursts, Blue Stars-Starbursts, Red  
Triangles-\ulg,
Green Circles-AGN \ulg, Purple Cross-\vv
\label{div_L_K}}
\end{figure}

\begin{figure}
\centering
\includegraphics[width=2in,angle=-90]{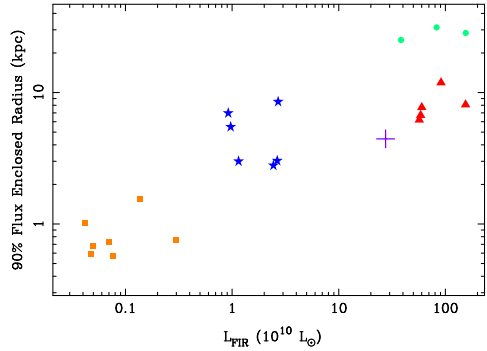}
\vspace{.2in}
\caption{
90\% Flux Enclosed Radii  in the 0.3-1.0 keV band vs % K-Band and 
FIR luminosities (see \citet{grim05} for details).
The size of the X-ray emitting region is correlated with the 
SFR rate as measured by the FIR luminosity.  
\vv~falls on the lower edge of the relation possibly 
as the soft X-ray band is only detectable in \vv W while
the FIR luminosity traces the SFR of the entire galaxy.
The displayed FIR luminosities of the dwarf starbursts are actually the sum of their UV and FIR luminosities.
We use their bolometric luminosity to account for the lower dust content in dwarfs.
Key: Orange Squares-Dwarf Starbursts, Blue Stars-Starbursts,
Red Triangles-\ulg, Green Circles-AGN \ulg, Purple Cross-\vv
\label{r90}}
\end{figure}

In our study we saw that the luminosity of the diffuse soft X-ray emission was 
roughly linearly proportional to the SFR (e.g. \lfir).  \vv~
is no exception as can be seen in Figure \ref{div_L_K}.  In this figure
we have divided both the FIR and X-ray luminosities by the K-Band luminosity.
By dividing by the K-band luminosity, a proxy for stellar mass,
we show that it not merely a matter of more massive galaxies having
higher SFRs and X-ray luminosities.  The linear scaling between
X-ray luminosity and \lfir ~suggests a constant efficiency in converting
mechanical energy from the galactic winds into emission from hot gas.  
The size scale of the region of X-ray emission will also depend upon the rate at which the starburst supplies mass and energy to the wind, and this relation is shown in Figure \ref{r90} where we have plotted the
90\% X-ray flux (0.3-1.0 keV) enclosed radius vs the FIR luminosity. Again, 
\vv ~follows the same trends seen in the other starbursts.

\begin{figure}
\centering
\includegraphics[width=2in,angle=-90]{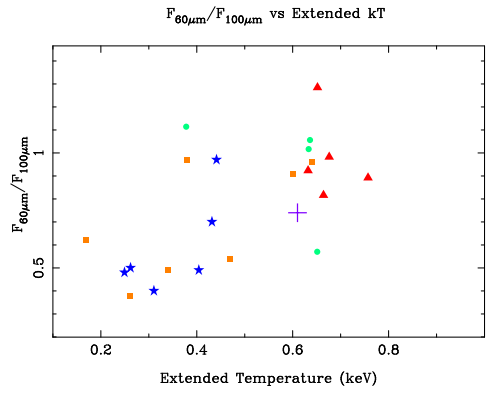}
\vspace{.2in}
\caption{
$\rm{F_{60\mu m}/F_{100\mu m}}$ vs  X-ray Gas Temperature.
The $\rm{F_{60\mu m}/F_{100\mu m}}$ ratio is an indicator of
dust temperature and hence the star formation rate per unit area. 
There is a correlation between gas
and dust temperature.  \vv ~ follows the same correlation as the starbursts.  
Key: Orange Squares-Dwarf Starbursts, Blue Stars-Starbursts,
Red Triangles-\ulg, Green Circles-AGN \ulg, Purple Cross-\vv
\label{F60o100vkt}}
\end{figure}

In \citet{grim05} we also showed there was a trend for the starbursts with the highest SFR per unit area (as traced by the FIR color-temperature of the warm dust) to have the hottest X-ray gas. Again \vv ~is consistent with this trend, as 
shown in Figure \ref{F60o100vkt}. Lastly, in agreement with the 
starbursts in \cite{grim05}, \vv ~appears to have an \afe~
element ratio that is enhanced (by a factor of $\sim$2) relative to solar abundances.  An
enhanced \afe~element ratio would be expected in a starburst
galaxy as the hot gas could be significantly enriched by the material injected by the core-collapse 
supernovae driving the winds (e.g. \citet{marc05}).

It is interesting to compare \vv~to the \ulg.  Although it  
follows the same scaling relations as the \ulg ~it does not appear 
to live in the same parameter space.  Obviously, it has 
a lower FIR luminosity and thus SFR. It also has a slightly smaller but similar K-Band
luminosity than the average \ulg ~(and hence a correspondingly similar stellar mass). Lastly \vv ~has slightly lower gas and dust temperatures than the
\ulg.  This suggests that \vv ~ is a similar mass galaxy
with a lower SFR per unit area than the \ulg ~and is intermediate between 
the normal starbursts and \ulg.

\subsection{Escape of Ionizing Radiation}

One of the major puzzles in cosmology is the nature of the objects responsible for the reionization of the universe, which may have started as early as z $\sim$30 and was complete by z$\sim$6 \citep{loeb01}. The known population of AGN do not appear to be sufficient by a wide margin, and the leading candidates are star forming galaxies \citep{sti04,pan05}.  \citet{steid01} reported the detection of a significant flux of ionizing radiation escaping from LBGs based on the detection of flux in the rest-frame Lyman continuum. However, other studies of high-z star forming galaxies have reached different conclusions \citep[e.g. ][]{malk03}.

There have been several attempts to measure the fraction of the ionizing radiation escaping from local starbursts ($f_{esc}$). The hope is that investigations of local
star forming galaxies will allow us to understand the physical processes that
determine $f_{esc}$
so that we can apply these lessons to high redshift galaxies for which our 
information is less complete.

\citet{leith95} reported the first direct measurements of
$f_{esc}$ using the Hopkins Ultraviolet Telescope to observe
below the rest-frame Lyman edge in a sample of four local starbursts,
and these data were later reanalyzed by
\citet{hurw97}. The
resulting upper limits on $f_{esc}$ were typically 10\%.  \citet{deh01}
 have obtained similar data with FUSE for the
starburst galaxy Mrk~54 at z = 0.0448. No flux was detected below the Lyman edge in the rest frame.
By comparison with the number of ionizing photons derived from the H$\alpha$
line, they set an upper limit to $f_{esc}$ of 6\%.

\citet{heck01b} used FUSE in a different way to constrain $f_{esc}$
in a sample five of the UV-brightest local starburst galaxies.
They showed that the strong \ion{C}{2}~$\lambda$1036 interstellar absorption-line is black in its core. Since the photoelectric opacity of the neutral ISM
below the Lyman-edge will be significantly larger than in the \ion{C}{2} line,
they were able to use these data to set a typical upper limit
on $f_{esc}$ of 6\% in these galaxies. Inclusion of absorption of Lyman
continuum photons by dust grains
will further decrease $f_{esc}$ (by up to an order of magnitude in some cases).
They also assessed the idea that
strong galactic winds can clear channels
through the neutral ISM of the galaxy and thereby increase $f_{esc}$
\citep[e.g. ][]{fuj03}.

Given the presence of a strong wind in \vv ~ and the overall similarity between
\vv ~and typical LBGs, it is interesting to determine $f_{esc}$ in this
object. As can be seen in Figures 2 and 3, the \ion{C}{2}~$\lambda$1036
line in \vv ~is quite black at line center.  This is particularly striking
as the large FUSE aperture samples all of the sightlines
towards \vv W.  Following the precepts
of \citet{heck01b} we find that $f_{esc}$ is smaller than a few
percent in \vv W.  The escape fraction decreases even further if we consider that all of the FUV radiation
from \vv E is also being absorbed. 

In this aspect then, \vv ~ is apparently quite different from
the LBGs studied by \citet{steid01}. It also differs from Haro~11, which is the only local case known
in which ionizing radiation is escaping from a galaxy \citep{berg06}. Since Haro~11 and \vv~are similar in
many ways, it will important to understand why they are different in this crucial respect.

\section{Conclusions}
Understanding the role played by the feedback from star formation in the evolution of galaxies and the IGM is a crucial problem in cosmology \citep{fuj04,scann05}. In the local universe, detailed observations of starburst galaxies have shown that this feedback is manifested most dramatically in the form of galactic winds that are driven by the collective effect of the kinetic energy supplied by winds from massive stars and supernova explosions. The outflows are complex multi-phase phenomena whose physical, chemical, and dynamical properties can only be understood through complemetary observations at many wavebands \citep{veill05}. Observations of the coronal ($10^5$ to $10^6$ K) and hot ($10^6$ to $10^7$\,K) gas are particularly important, as they provide essential information about the importance of radiative cooling of the outflow and about the dynamics and energy content of the wind. The coronal gas is best traced through the \ion{O}{6} doublet in the far-UV and the hot g!
as is best traced via its soft X-ray emission.

The strong overall cosmic evolution in the global star formation rate  \citep{bunk04} means that the bulk of the feedback from galactic winds occurred at early times ($z >$ 1). Indeed, the direct signature of galactic winds -- the presence of broad, blueshifted interstellar absorption lines in the rest-frame UV -- is generically present in the Lyman Break Galaxies \citep[LBGs, ][]{shap03}. These are the best-studied population of high-redshift star forming galaxies  \citep{steid99}. However, such data provide only a narrow range of information about galactic winds. Unfortunately, direct observations of the hotter gas in the rest-frame soft X-ray and FUV regions for high redshift galaxies are extremely difficult or even impossible, which hinders progress in understanding 
this complex phenomenon.  Finding and studying nearby analogs to LBGs is therefore very important.

In this paper we have described FUV and soft X-ray observations of \vv, the nearest known galaxy whose basic properties are a good match to those of typical LBGs 
\citep{heck05,hoop06}. Our \fu~observations in the FUV show strong and very broad interstellar absorption lines. The lines with the highest S/N show two kinematic components. One is centered near the galaxy systemic velocity and apparently corresponds to the galaxy ISM. The second component is strongly blueshifted, with a centroid  300 to 400 $\ks$ below systemic velocity and weaker absorption extending to blueshifts as high as $\sim1000~ \ks$. This is consistent with what is seen at slightly longer rest wavelengths
in high redshift LBGs and further establishes the similarity of \vv~to the LBGs.  
The \ovi ~absorption line covers a similar range in velocity to the other ions, but has a much more symmetric profile. This suggests a different origin for the coronal-phase absorbing gas.  The high column density and broad width of the
\ovi~line are consistent with production of \ovi~in gas that is cooling radiatively from high temperature through the coronal regime. One possible origin for this gas is at the interface between the hot outflowing gas seen in X-rays and the cool clouds in the galaxy halo.
These cooler clouds may be traced by the absorption lines from the cooler gas seen in the \fu~ observations.

Observations with \xmm~and \ch~of the hot diffuse gas 
in \vv~are consistent with those seen in a local sample of star forming galaxies with galactic outflows. As expected based on its far-IR luminosity and implied SFR, \vv ~has X-ray properties intermediate between those of \ulg ~and those of more typical present-day starbursts. This suggests
that diffuse thermal X-ray emission should be a common feature of LBGs created in
the shocks between the outflowing wind material and surrounding medium \citep{marc05}.
Hard X-ray, far-IR, and radio continuum observations suggest that half of the star formation in \vv~is not easily observable
in the UV and soft X-ray as it is taking place in the heavily obscured eastern component of
\vv.

The strong wind in \vv ~ might in principle be able to carve a channel in the ISM through
which ionizing photons could escape from the starburst to the IGM. However,
we show that the fact that the core of the \ion{C}{2}~$\lambda$1036 
absorption line is black implies that the fraction of ionizing photons that
escape from \vv ~ is no more than a few percent. This low escape fraction 
in \vv ~ is quite different from that seen in the LBGs studied by \citet{steid01}.

As the sample of nearby LGB-analogs discovered by GALEX grows, we will be able to eventually conduct multi-waveband investigations of the galactic winds in a sample large enough for us to be able to make robust statements about the physical, chemical, and dynamical properties of these objects. This should have important implications for the evolution of galaxies and the IGM.

%\newpage

\onecolumn

%\newpage
\clearpage
\begin{deluxetable}{ccccccccccc}
\tabletypesize{\scriptsize}
\tablecolumns{11}
\tablecaption{Basic Properties of \vv \label{prop}}
\tablehead{
\colhead{} & \multicolumn{2}{c}{Position} & \colhead{cz\tablenotemark{a}} & \colhead{Scale} & 
\colhead{Z\tablenotemark{b}} &
\colhead{$\rm{L_{FIR}}$\tablenotemark{c}} & \colhead{$\rm{L_{K}}$\tablenotemark{c}} & \colhead{$\rm{L_{FUV}}$\tablenotemark{d}} & \colhead{$\rm{r_{50,UV}}$\tablenotemark{d}} &
\colhead{$\rm{I_{FUV}}$\tablenotemark{c}} \\
\colhead{} & \multicolumn{2}{c}{J2000} &  \colhead{$\ks$} & \colhead{kpc/\arcsec} & &
\colhead{$\rm{L_\odot}$} & \colhead{$\rm{L_\odot}$} & \colhead{$\rm{L_\odot}$}  & 
\colhead{kpc} & \colhead{$\rm{L_\odot/kpc^2}$} }
\startdata
Global & +1 07 47.1 & -17 30 24.3 & 6040 & 0.41 & 8.6-8.7 & $2.8\times10^{11}$ & $2.1\times10^{10}$ & $3.2\times10^{10}$\\
East & +1 07 47.6 & -17 30 25.8 & & & & & & $1.0\times10^{10}$ & 4 & $9.9\times10^7$\\
West & +1 07 46.6 & -17 30 24.0 & & & & & & $2.2\times10^{10}$ & 2.3 & $5.6\times10^8$\\
\enddata
\tablenotetext{a}{\citet{yun94}.}
\tablenotetext{b}{Gas phase metallicity as log(O/H)+12 from \citet{kim95} using the transformations of \citet{char01}.}
\tablenotetext{c}{IRAS \citep{soif89}  and 2Mass \citep{jarr03} results are transformed using methods of \citet{sand96} and \citet{carp01} respectively.}
\tablenotetext{d}{Results derived from \citet{gold02}.}
\end{deluxetable}

\begin{deluxetable}{lcccccc}
\tabletypesize{\footnotesize}
\tablecolumns{7}
\tablewidth{0pc}
\tablecaption{Single Component ISM Absorption Line Fit Data}
\tablehead{
\colhead{Ion} & \colhead{$\lambda_{0}$} & \colhead{$log(\lambda f N/N_H)$} &
                 \colhead{Instrument} & \colhead{$\rm{W}_\lambda$} & 
                 \colhead{$v_{c}$} & \colhead{FWHM} \\
                  & \colhead{$\ang$} & & & \colhead{$\ang$} & \colhead{km/s} & 
                  \colhead{km/s}\label{absdata}}
\startdata
Ly~$\beta$ & 1025.722 & 1.909 & lif1a &  $3.8\pm0.5$ & $5975\pm24$ & $728$\tablenotemark{a}\\
 & & & lif2b &  $3.9\pm0.6$ & $5962\pm34$ & $728$\tablenotemark{a}\\
\ion{C}{3} & 977.02 & -0.608 & lif1a &  $3.2\pm0.8$ & 5915\tablenotemark{b} & $700$\tablenotemark{b}\\
 & & & lif2b &  $3.6\pm1.4$ & $5929\pm59$ & $700$\tablenotemark{b}\\
 \ovi & 1031.926 & -1.133 & lif1a &  $1.8\pm0.1$ & 5771\tablenotemark{a} & $700$\tablenotemark{b}\\
 & & & lif2b &  $2.0\pm0.1$ & 5845\tablenotemark{a} & $700$\tablenotemark{b}\\
\ion{C}{2} & 1036.337 & -1.374 & lif1a &  $2.8\pm0.2$ & $6011\pm17$ & $655\pm59$\\
 & & & lif2b &  $2.8\pm0.2$ & $5978\pm24$ & $658\pm80$\\
\ion{O}{1} & 988.733 & -1.564 & lif1a &  $1.6\pm0.4$ & $5935\pm30$ & $700$\tablenotemark{b}\\
 & & & lif2b &  $2.2\pm0.4$ & 5932\tablenotemark{a} & $700$\tablenotemark{b}\\
\ion{N}{2} & 1083.99 & -2.002 & lif1b &  $2.5\pm0.1$ & $5954\pm22$ & $693$\tablenotemark{a}\\
 & & & lif2a &  $2.6\pm0.1$ & $5949\pm19$ & $683$\tablenotemark{a}\\
\ion{N}{3} & 989.799 & -2.027 &  lif1a & $2.8\pm0.2$ & 5935\tablenotemark{a} & $700$\tablenotemark{b}\\
 & & & lif2b &  $1.8\pm0.4$ & 5932\tablenotemark{c} & $700$\tablenotemark{b}\\
\ion{N}{1} & 1134.415 & -2.089 & lif1b &  $1.0\pm0.2$ & $5917\pm47$ & $700$\tablenotemark{b}\\
 & & & lif2a &  $1.1\pm0.2$ & $5918\pm34$ & $656\pm116$\\
\ion{O}{1} & 1039.23 & -2.290 & lif1a &   $0.8\pm0.2$ & $6011$\tablenotemark{d} & $655$\tablenotemark{d}\\
 & & & lif2b &  $0.8\pm0.2$ & $5978$\tablenotemark{d} & $658$\tablenotemark{d}\\
 \ion{Fe}{2}\tablenotemark{d} & 1144.938 & -2.420 & lif1b &  $1.3\pm0.1$ & 5698\tablenotemark{a} & $700$\tablenotemark{b}\\
 & & & lif2a &  $1.1\pm0.1$ & 5790\tablenotemark{a} & $700$\tablenotemark{b}\\
% \ion{S}{3} & 1012.502 & -3.244 & lif1a &  $1.4\pm0.1$ & $5816\pm36$ & $639\pm70$\\
% & & & lif2b &  $1.2\pm0.2$ & 5814\tablenotemark{a} & $700$\tablenotemark{b}\\
% \ion{S}{4} & 1073.2 & lif2a & $0.4\pm0.1$ & $5810\pm18$ & $390$\tablenotemark{c}\\
\enddata
\tablenotetext{a}{Unable to determine errors on this value.}
\tablenotetext{b}{Unable to accurately determine value so FWHM fixed to 700~$\ks$}
\tablenotetext{d}{\ion{O}{1}~$\lambda$1039 values tied to velocity shift and FWHM of the \ion{C}{2}~$\lambda$1036 line.}
\tablenotetext{c}{\ion{N}{3}~$\lambda$990 values tied to velocity shift of the \ion{O}{1}~$\lambda$989 line.}
\tablenotetext{d}{\ion{Fe}{2}~$\lambda$1144 line measurements are strongly affected by the second order airglow emission line \ion{He}{1}~$\lambda$584 
and an instrumental artifact that causes a break in the continuum}
\end{deluxetable}

\begin{deluxetable}{lccccccccccc}
\rotate 
\tabletypesize{\scriptsize}
\tablecolumns{12}
\tablewidth{0pc}
\tablecaption{Two Component Absorption Line Fits}
\tablehead{
& & & & & \multicolumn{3}{c}{Galaxy} & & \multicolumn{3}{c}{Outflow} \\
\cline{6-8} \cline{10-12}
\colhead{Ion} & \colhead{$\lambda_{0}$} & \colhead{$log(\lambda f N/N_H)$} &
                 \colhead{Instrument} & \phantom{ } & \colhead{$\rm{W}_\lambda$} & 
                 \colhead{$v_{c}$} & \colhead{FWHM} & \phantom{ } & \colhead{$\rm{W}_\lambda$} & 
                 \colhead{$v_{c}$} & \colhead{FWHM} \\
                  & \colhead{$\rm{\AA}$} & & & & \colhead{$\rm{\AA}$} & \colhead{km/s} & 
                  \colhead{km/s} & & \colhead{$\rm{\AA}$} & \colhead{km/s} & 
                  \colhead{km/s}\label{twoabsdata}}
\startdata
\ion{C}{2} & 1036.337 & -1.374 & lif1a &  & $2.2\pm0.4$ & $6081\pm32$ & $429\pm85$ & & $0.9\pm0.5$ & $5663\pm126$ & $420\pm140$\\
 & & & lif2b & &$2.2\pm0.6$ & $6079\pm17$ & $411\pm143$ & & $1.3\pm0.5$ & $5697\pm97$ & $503\pm176$\\ 
\ion{N}{2} & 1083.99 & -2.002 & lif1b &  & $1.8\pm0.1$ & $6019\pm11$ & $499\pm49$ & & $0.7\pm0.3$ & $5623\pm117$ & $349$\tablenotemark{a}\\
 & & & lif2a & & $2.0\pm0.1$ & $6008\pm31$ & $529\pm57$ & & $0.8\pm0.4$ & $5681\pm178$ & $408$\tablenotemark{a}\\
\ion{N}{1} & 1134.415 & -2.089 & lif1b &  & $0.6\pm0.2$ & $6076\pm45$ & $318\pm90$ & & $0.4\pm0.2$ & $5669\pm73$ & $302\pm145$\\
 & & & lif2a & &$0.7\pm0.2$ & $6035\pm49$ & $372\pm101$ & & $0.3\pm0.2$ & $5669\pm60$ & $270\pm165$\\
\ion{O}{1} & 1039.23 & -2.290 & lif1a & &$0.7\pm0.2$ & $6081$\tablenotemark{b} & $429$\tablenotemark{b} & & $0.3\pm0.2$\tablenotemark{b} & $5663$\tablenotemark{b} & $420$\tablenotemark{b}\\
 & & & lif2b & &$0.7\pm0.3$ & $6079$\tablenotemark{b} & $411$\tablenotemark{b} & & $0.2\pm0.4$ & $5697$\tablenotemark{b} & $503$\tablenotemark{b}\\ 
\enddata
\tablenotetext{a}{Unable to determine errors on this value.}
\tablenotetext{b}{\ion{O}{1}~$\lambda$1039 values tied to equivalent value of the \ion{C}{2}~$\lambda$1036 line.}
\end{deluxetable}

\begin{deluxetable}{lccccc}
\tabletypesize{\footnotesize}
\tablecolumns{6}
\tablewidth{0pc}
\tablecaption{Stellar Photospheric Absorption Line Fit Data}
\tablehead{
\colhead{Ion} & \colhead{$\lambda_{0}$} &
                 \colhead{Instrument} & \colhead{$\rm{W}_\lambda$} & 
                 \colhead{$v_{c}$} & \colhead{FWHM} \\
                  & \colhead{$\ang$} & & \colhead{$\ang$} & \colhead{km/s} & 
                  \colhead{km/s}\label{phabsdata}}
\startdata
\ion{Si}{4} & 1122.487 & lif1b & $1.1\pm0.1$ & $5967\pm22$ & $390$\tablenotemark{a}\\
 & & lif2a & $1.0\pm0.1$& $5974\pm18$ &  $374\pm42$\\
\ion{Si}{4} & 1128.201 & lif1b & $0.8\pm0.1$ & $5967$\tablenotemark{b} & $390$\tablenotemark{b}\\
 & & lif2a & $0.7\pm0.1$ & $5974$\tablenotemark{b} &  $389$\tablenotemark{a}\\
 \ion{P}{5} & 1117.977 & lif1b & $0.3\pm0.1$ & $5960$\tablenotemark{a} & $390$\tablenotemark{c}\\
 & & lif2a & $0.3\pm0.1$ & $5950$\tablenotemark{a} &  $390$\tablenotemark{c}\\
\enddata
\tablenotetext{a}{Unable to determine errors on this value.}
\tablenotetext{b}{Value tied to velocity shift or FWHM of the \ion{Si}{4}~$\lambda$1122 line.}
\tablenotetext{c}{Unable to accurately determine value so FWHM fixed to 390~$\ks$}
\end{deluxetable}

\begin{deluxetable}{lccccccclcccc}
\rotate
\tabletypesize{\scriptsize}
\tablecolumns{13}
\tablewidth{0pc}
\tablecaption{X-Ray Spectral Fits}
\tablehead{
%\colhead{wabs1} & \multicolumn{3}{c}{vapec} & \colhead{wabs2} & \multicolumn{2}{c}{powerlaw} & \\
\colhead{Region} & \colhead{Instr.} & \multicolumn{2}{c}{Position\tablenotemark{a}} & \colhead{Radius\tablenotemark{a}} & \colhead{$\rm{N_H}$} & \colhead{kT}   & \colhead{$K$\tablenotemark{b}} & \colhead{$\alpha/\rm{Fe}$} & 
\colhead{$N_{H}$} & \colhead{PL Norm\tablenotemark{c}}   & \colhead{$\Gamma$} & 
\colhead{$\chi^2/\rm{DOF}$}\\
& & \multicolumn{2}{c}{J2000} & \colhead{\arcsec} & \colhead{$10^{22}~\rm{cm^{-2}}$} & \colhead{keV}    & \colhead{} & \colhead{} &
\colhead{$10^{22}~\rm{cm^{-2}}$}   & \colhead{}    & \colhead{} \label{xspecdata}}
\startdata
\\
\vv\tablenotemark{d} & Chandra & 01:07:47.0 & -17 30 29.0 &  $24.3$ & $5.7\times10^{-2}$ & $0.62^{+0.03}_{-0.03}$ & $2.6^{+0.7}_{-1.0}\times10^{-4}$ & $2.7^{+0.6}_{-0.7}$ & 
$0.6^{+0.3}_{-0.3}$ & $9.2^{+4}_{-3}\times10^{-5}$ & $2.0^{+0.3}_{-0.2}$ & 121/128\\
\vv\tablenotemark{d} & XMM       & 01:07:47.0 & -17 30 29.0 &  $24.3$ & $5.7\times10^{-2}$ & $0.59^{+0.06}_{-0.07}$ & $2.5^{+2.5}_{-2.5}\times10^{-4}$ & $2.2^{+2.00}_{-0.73}$ & 
$1.1^{+1.1}_{-0.7}$ & $2.2^{+4.5}_{-1.1}\times10^{-4}$ & $2.8^{+1.2}_{-1.5}$ & 100/117\\
\\
\cline{1-13} 
\\
\vv E\tablenotemark{e} & Chandra & 01 07 47.5 & -17 30 25.6 & 4.4  & $5.7\times10^{-2}$ & $0.30^{+0.75}_{-0.2}$
& $1.7^{+6.0}_{-1.4}\times10^{-4}$ & $5.0\tablenotemark{f}$ &  $1.5^{+3.4}_{-1.0}$ & $1.7^{+2.2}_{-1.1}\times10^{-5}$ & $1.3^{+0.2}_{-0.5}$ & 13/16
\\ \\
\enddata
\tablenotetext{a}{Center and Radius of circular region used to extract fitted spectra.}
\tablenotetext{b}{Plasma model normalization in units of 
$\frac{10^{-14}}{4\pi [D_A (1+z)]^2}\int n_e n_H dV$, where $D_A$ is the angular distance, and $n_H$ and $n_e$ are the hydrogen and electron number densities respectively.}
\tablenotetext{c}{$\rm{photons~keV^{-1}\,cm^{-2}\, s^{-1}}$ at 1 keV}
\tablenotetext{d}{Values derived from an xspec model of  wabs1 ( vmekal + zwabs2 ( powerlaw )) using default abundances
angr \citep{and89}.}
\tablenotetext{e}{Values derived from an xspec model of  wabs1 ( zwabs2 ( vmekal + gaussian1 + gaussian2) + zwabs3 ( powerlaw )) using default abundances
angr \citep{and89}.  The gaussians were centered at 1.83 keV and 1.39 keV with equivalent widths of 1.67 keV and .31 keV respectively.  The derived value for the other intrinsic column absorber for the thermal component and line emission was  $\rm{N_H\sim8\times10^{21}\,cm^{-2}}$. }  
\tablenotetext{f}{As we observe photoionized emission in \vv E it is not clear that the \afe~ratio of the vmekal model is physically meaningful. }
\end{deluxetable}
%01:07:46.959 -17:30:29.03 24.2728
%01:07:47.535 -17:30:25.59 4.39977

\begin{deluxetable}{lcccccc}
\tabletypesize{\footnotesize}
\tablecolumns{7}
\tablewidth{0pc}
\tablecaption{Mass and Energy Outflow Rates}
\tablehead{
\colhead{Ion}  &  \colhead{$\rm{N_{Ion}}$\tablenotemark{a}} & 
\colhead{Abundance\tablenotemark{b}} & \colhead{$\rm{N_H}$}   & 
\colhead{$\frac{\Omega}{4\pi}\,\dot{\rm{M}}$\tablenotemark{c}} & 
\colhead{$\frac{\Omega}{4\pi}\dot{\rm{E}}$\tablenotemark{c}}\\
\colhead{} & \colhead{$10^{14}~\rm{cm^{-2}}$}   & \colhead{$\rm{\log(N_{Elem}/N_H)+12}$}    & \colhead{$10^{18}~\rm{cm^{-2}}$} & 
\colhead{$\rm{M_\odot/yr}$} & \colhead{$\rm{10^{40}~ergs/s}$}\label{outflow}}
\startdata
\ion{O}{6} & 21 & 8.66 & $>$23 & $>$2.6 & $>$13 \\
\ion{N}{2} & 8.4 & 7.78 & 14 & 1.54 & 7.8 \\
\ion{N}{1} & 4.3 & 7.78 & 7.1 & 0.79 & 4.0 \\
\ion{N}{1}+\ion{N}{2} & 13 & 7.78 & 21 & 2.3 & 12\\
\ion{C}{2} & 13 & 8.39 & 5.1 & 0.56 & 2.9\\
Total Cool Gas\tablenotemark{d}       & & & $>$44 & $>$4.9 & $>$25\\
\cline{1-6} 
Total Hot Gas\tablenotemark{e}       & & &  & $50f^{1/2}$ & $500f^{1/2}$\\ 
\enddata
\tablenotetext{a}{Based on outflow equivalent width and FWHM from Table \ref{twoabsdata}.}
\tablenotetext{b}{\citet{asp05}}
\tablenotetext{c}{Estimated mass and kinetic energy outflow rates (see text).} 
\tablenotetext{d}{From FUSE data} 
\tablenotetext{e}{From X-ray data.  Note that f is the volume filling factor of the hot gas.} 
\end{deluxetable}

\clearpage

\clearpage

\begin{thebibliography}{}

\bibitem[Aguirre et al.(2005)]{aguir05} Aguirre, A., Schaye, 
J., Hernquist, L., Kay, S., Springel, V., \& Theuns, T.\ 2005, \apjl, 620, 
L13 

\bibitem[Anders \& Grevesse(1989)]{and89} Anders, E., \& 
Grevesse, N.\ 1989, \gca, 53, 197

\bibitem[Asplund et al.(2005)]{asp05} Asplund, M., Grevesse, 
N., \& Sauval, A.~J.\ 2005, ASP Conf.~Ser.~336: Cosmic Abundances as 
Records of Stellar Evolution and Nucleosynthesis, 336, 25 

\bibitem[Bauer et al.(2002)]{bauer02} Bauer, F.~E., Alexander, 
D.~M., Brandt, W.~N., Hornschemeier, A.~E., Vignali, C., Garmire, G.~P., \& 
Schneider, D.~P.\ 2002, \aj, 124, 2351 

\bibitem[Bennett et al.(2003)]{benn03} Bennett, C.~L., et al.\ 
2003, \apjs, 148, 1 

\bibitem[Bergvall et al.(2006)]{berg06} Bergvall, N., Zackrisson, E., Andersson, B.-G., Arnberg, D., Masegoas, J., \& Ostlin, G.\
2006, A\&A, 448, 513

\bibitem[Bunker et al.(2004)]{bunk04} Bunker, A.~J., Stanway, 
E.~R., Ellis, R.~S., \& McMahon, R.~G.\ 2004, \mnras, 355, 374 

\bibitem[Carpenter(2001)]{carp01} Carpenter, J.~M.\ 2001, \aj,
121, 2851   

\bibitem[Chandra Interactive Analysis of Observations (CIAO)]{ciao}
Chandra Interactive Analysis of Observations (CIAO), http://cxc.harvard.edu/ciao/

\bibitem[Charlot \& Longhetti(2001)]{char01} Charlot, S., \& 
Longhetti, M.\ 2001, \mnras, 323, 887 

\bibitem[Colbert et al.(2004)]{colb04} Colbert, E.~J.~M., 
Heckman, T.~M., Ptak, A.~F., Strickland, D.~K., \& Weaver, K.~A.\ 2004, 
\apj, 602, 231 

\bibitem[Condon et al.(1991)]{con91} Condon, J.~J., Huang, 
Z.-P., Yin, Q.~F., \& Thuan, T.~X.\ 1991, \apj, 378, 65 

\bibitem[Davis et al.(1985)]{davis85} Davis, M., Efstathiou, 
G., Frenk, C.~S., \& White, S.~D.~M.\ 1985, \apj, 292, 371 

\bibitem[Deharveng et al.(2001)]{deh01} Deharveng, J.-M., 
Buat, V., Le Brun, V., Milliard, B., Kunth, D., Shull, J.~M., \& Gry, C.\ 
2001, \aap, 375, 805 

\bibitem[Dixon \& Sahnow(2003)]{dix03} Dixon, W.~V., \& 
Sahnow, D.~J.\ 2003, ASP Conf.~Ser.~295: Astronomical Data Analysis 
Software and Systems XII, 295, 241 

\bibitem[Dove et al.(2000)]{dov00} Dove, J.~B., Shull, J.~M., 
\& Ferrara, A.\ 2000, \apj, 531, 846 

\bibitem[Doyon et al.(1995)]{doy95} Doyon, R., Nadeau, D., 
Joseph, R.~D., Goldader, J.~D., Sanders, D.~B., \& Rowlands, N.\ 1995, 
\apj, 450, 111


\bibitem[Le Floc'h et al.(2002)]{floc02} Le Floc'h, E., 
Charmandaris, V., Laurent, O., Mirabel, I.~F., Gallais, P., Sauvage, M., 
Vigroux, L., \& Cesarsky, C.\ 2002, \aap, 391, 417 

\bibitem[Franceschini et al.(2003)]{franc03} Franceschini, A., 
et al.\ 2003, \mnras, 343, 1181 

\bibitem[Fujita et al.(2003)]{fuj03} Fujita, A., Martin, 
C.~L., Mac Low, M.-M., \& Abel, T.\ 2003, \apj, 599, 50

\bibitem[Fujita et al.(2004)]{fuj04} Fujita, A., Mac Low, 
M.-M., Ferrara, A., \& Meiksin, A.\ 2004, \apj, 613, 159 

\bibitem[Giavalisco(2002)]{giav02} Giavalisco, M.\ 2002, 
\araa, 40, 579 

\bibitem[Gilfanov et al.(2004)]{gilf04} Gilfanov, M., Grimm, 
H.-J., \& Sunyaev, R.\ 2004, Nuclear Physics B Proceedings Supplements, 
132, 369 

\bibitem[Goldader et al.(2002)]{gold02} Goldader, J.~D., 
Meurer, G., Heckman, T.~M., Seibert, M., Sanders, D.~B., Calzetti, D., \& 
Steidel, C.~C.\ 2002, \apj, 568, 651 

\bibitem[Grimes et al.(2005)]{grim05} Grimes, J.~P., Heckman, 
T., Strickland, D., \& Ptak, A.\ 2005, \apj, 628, 187 

\bibitem[Heckman, Armus, \& Miley(1990)]{heck90} Heckman,
T.~M., Armus, L., \& Miley, G.~K.\ 1990, \apjs, 74, 833

\bibitem[Heckman et al.(1998)]{heck98} Heckman, T.~M., Robert,
C., Leitherer, C., Garnett, D.~R., \& van der Rydt, F.\ 1998, \apj, 503,
646

\bibitem[Heckman et al.(2000)]{heck00} Heckman, T.~M., Lehnert, M., Strickland, D., \& Armus, L.\ 2000,
\apjs, 129, 493

\bibitem[Heckman et al.(2001a)]{heck01} Heckman, T.~M., 
Sembach, K.~R., Meurer, G.~R., Strickland, D.~K., Martin, C.~L., Calzetti, 
D., \& Leitherer, C.\ 2001, \apj, 554, 1021 

\bibitem[Heckman et al.(2001b)]{heck01b} Heckman, T.~M., 
Sembach, K.~R., Meurer, G.~R., Leitherer, C., Calzetti, D., \& Martin, 
C.~L.\ 2001, \apj, 558, 56 

\bibitem[Heckman et al.(2002)]{heck02} Heckman, T.~M., Norman, 
C.~A., Strickland, D.~K., \& Sembach, K.~R.\ 2002, \apj, 577, 691

\bibitem[Heckman(2003)]{heck03} Heckman, T.~M.\ 2003, Revista 
Mexicana de Astronomia y Astrofisica Conference Series, 17, 47 

\bibitem[Heckman et al.(2005)]{heck05} Heckman, T.~M., et al.\ 
2005, \apjl, 619, L35 

\bibitem[Helsdon \& Ponman(2000)]{hels00} Helsdon, S.~F., \& 
Ponman, T.~J.\ 2000, \mnras, 315, 356 

\bibitem[Hoopes, Heckman, Strickland, \& Howk(2003)]{hoop03}
Hoopes, C.~G., Heckman, T.~M., Strickland, D.~K., \& Howk, J.~C.\ 2003,
\apjl, 596, L175

\bibitem[Hoopes et al.(2006)]{hoop06}
Hoopes, C.~G., et al. \ 2006, in prep

\bibitem[Hurwitz et al.(1997)]{hurw97} Hurwitz, M., Jelinsky, 
P., \& Dixon, W.~V.~D.\ 1997, \apjl, 481, L31

\bibitem[Iono et al.(2004)]{iono04} Iono, D., Ho, P.~T.~P., 
Yun, M.~S., Matsushita, S., Peck, A.~B., \& Sakamoto, K.\ 2004, \apjl, 616, 
L63 

\bibitem[Jarrett et al.(2003)]{jarr03} Jarrett, T.~H.,
Chester, T., Cutri, R., Schneider, S.~E., \& Huchra, J.~P.\ 2003, \aj,
125, 525

\bibitem[Kennicutt(1998)]{kenn98} Kennicutt, R.~C.\ 1998, 
\apj, 498, 541 

\bibitem[Kim et al.(1995)]{kim95} Kim, D.-C., Sanders, D.~B., 
Veilleux, S., Mazzarella, J.~M., \& Soifer, B.~T.\ 1995, \apjs, 98, 129

\bibitem[Klypin et al.(1999)]{klyp99} Klypin, A., Kravtsov, 
A.~V., Valenzuela, O., \& Prada, F.\ 1999, \apj, 522, 82 

\bibitem[Kriss(1994)]{kriss94} Kriss, G.\ 1994, ASP Conf.~Ser.~ 
61: Astronomical Data Analysis Software and Systems III, 61, 437

\bibitem[Lehmer et al.(2005)]{lehm05} Lehmer, B.~D., et al.\ 
2005, \aj, 129, 1 

\bibitem[Lehnert \& Heckman(1996)]{lehn96} Lehnert, M.~D.~\&
Heckman, T.~M.\ 1996, \apj, 462, 651

\bibitem[Levenson et al.(2006)]{lev06} Levenson, N.~A.,
Heckman, T.~M., Krolik, J.~H., Weaver, K.~A., \& Zycki, P.,~T. \ 2006, submitted

\bibitem[Leitherer et al.(1995)]{leith95} Leitherer, C., 
Ferguson, H.~C., Heckman, T.~M., \& Lowenthal, J.~D.\ 1995, \apjl, 454, L19 

\bibitem[Loeb \& Barkana(2001)]{loeb01} Loeb, A., \& Barkana, 
R.\ 2001, \araa, 39, 19 

\bibitem[Malkan et al.(2003)]{malk03} Malkan, M., Webb, W., \& 
Konopacky, Q.\ 2003, \apj, 598, 878 

\bibitem[Marcolini et al.(2005)]{marc05} Marcolini, A., 
Strickland, D.~K., D'Ercole, A., Heckman, T.~M., \& Hoopes, C.~G.\ 2005, 
\mnras, 362, 626 

\bibitem[Martin(2005)]{mart05b} Martin, C.~L.\ 2005, \apj, 621, 227

\bibitem[Martin et al.(2005)]{mart05} Martin, D.~C., et al.\ 
2005, \apjl, 619, L1 

\bibitem[Moos et al.(2000)]{moos00} Moos, H.~W., et al.\ 2000, 
\apjl, 538, L1 

\bibitem[Nandra et al.(2002)]{nan02} Nandra, K., Mushotzky, 
R.~F., Arnaud, K., Steidel, C.~C., Adelberger, K.~L., Gardner, J.~P., 
Teplitz, H.~I., \& Windhorst, R.~A.\ 2002, \apj, 576, 625 

\bibitem[Oskinova et al.(2006)]{osk06} Oskinova, L.~M.,
Feldmeier, A, Hamann, W.~R. (astro-ph 0603286)

\bibitem[Panagia et al.(2005)]{pan05} Panagia, N., Fall, 
S.~M., Mobasher, B., Dickinson, M., Ferguson, H.~C., Giavalisco, M., Stern, 
D., \& Wiklind, T.\ 2005, \apjl, 633, L1

\bibitem[Peacock et al.(2000)]{peac00} Peacock, J.~A., et al.\ 
2000, \mnras, 318, 535 

\bibitem[Pettini et al.(2002)]{pett02} Pettini, M., Rix, 
S.~A., Steidel, C.~C., Adelberger, K.~L., Hunt, M.~P., \& Shapley, A.~E.\ 
2002, \apj, 569, 742 

\bibitem[Robertson et al.(2005)]{rob05} Robertson, B., 
Bullock, J.~S., Font, A.~S., Johnston, K.~V., \& Hernquist, L.\ 2005, \apj, 
632, 872 

\bibitem[Rupke et al.(2005)]{rup05} Rupke, D.~S., Veilleux, S., \& Sanders, D.~B. 2005, \apjs,
160, 115

\bibitem[Sanders \& Mirabel (1996)]{sand96} Sanders, D.~B.~\&
Mirabel, I.~F.\ 1996, \araa, 34, 749  

\bibitem[Sako et al.(2002)]{sak02} Sako, M., Kahn, S.~M., 
Paerels, F., Liedahl, D.~A., Watanabe, S., Nagase, F., \& Takahashi, T.\ 
2002, High Resolution X-ray Spectroscopy with XMM-Newton and Chandra

\bibitem[Scannapieco et al.(2005)]{scann05} Scannapieco, C., 
Tissera, P.~B., White, S.~D.~M., \& Springel, V.\ 2005, \mnras, 907 

\bibitem[Schild et al.(2004)]{schil04} Schild, H., et al.\ 
2004, \aap, 422, 177 

\bibitem[Shapley et al.(2003)]{shap03} Shapley, A.~E., 
Steidel, C.~C., Pettini, M., \& Adelberger, K.~L.\ 2003, \apj, 588, 65 

\bibitem[Soifer, Boehmer, Neugebauer, \&
Sanders(1989)]{soif89} Soifer, B.~T., Boehmer, L., Neugebauer,
G., \& Sanders, D.~B.\ 1989, \aj, 98, 766

\bibitem[Sommer-Larsen et al.(1999)]{som99} Sommer-Larsen, 
J., Gelato, S., \& Vedel, H.\ 1999, \apj, 519, 501 

\bibitem[Springel et al.(2005)]{spring05} Springel, V., et al.\ 
2005, \nat, 435, 629 

\bibitem[Steidel et al.(1999)]{steid99} Steidel, C.~C., 
Adelberger, K.~L., Giavalisco, M., Dickinson, M., \& Pettini, M.\ 1999, 
\apj, 519, 1 

\bibitem[Steidel et al.(2001)]{steid01} Steidel, C.~C., 
Pettini, M., \& Adelberger, K.~L.\ 2001, \apj, 546, 665 

\bibitem[Stiavelli et al.(2004)]{sti04} Stiavelli, M., Fall, 
S.~M., \& Panagia, N.\ 2004, \apjl, 610, L1 

\bibitem[Strickland \& Stevens(2000)]{strick00} Strickland, 
D.~K., \& Stevens, I.~R.\ 2000, \mnras, 314, 511

\bibitem[Sutherland \& Dopita(1993)]{suth93} Sutherland, 
R.~S., \& Dopita, M.~A.\ 1993, \apjs, 88, 253 

\bibitem[Tamura et al.(2004)]{tam04} Tamura, T., Kaastra, 
J.~S., den Herder, J.~W.~A., Bleeker, J.~A.~M., \& Peterson, J.~R.\ 2004, 
\aap, 420, 135 

\bibitem[Tremonti et al.(2004)]{trem04} Tremonti, C.~A., et 
al.\ 2004, \apj, 613, 898 

\bibitem[Veilleux et al.(2005)]{veill05} Veilleux, S., Cecil, 
G., \& Bland-Hawthorn, J.\ 2005, \araa, 43, 769 

\bibitem[York et al.(2000)]{york00} York, D.~G., et al.\ 2000, 
\aj, 120, 1579 

\bibitem[Yun et al.(1994)]{yun94} Yun, M.~S., Scoville, 
N.~Z., \& Knop, R.~A.\ 1994, \apjl, 430, L109 

\end{thebibliography}
\end{document}